\documentclass[sn-chicago]{sn-jnl}% Chicago-based Humanities Reference Style
\jyear{2022}%

\theoremstyle{thmstyleone}%
\theoremstyle{thmstyletwo}%

\theoremstyle{thmstylethree}%
\newcounter{example}[section]

\newcommand\posscite[1]{\citeauthor{#1}'s (\citeyear{#1})}
\usepackage{capt-of}
\usepackage{xcolor}

\begin{document}
%new deadline: 31.1.23

\title[Evaluation of Embedding Models for NER]{Embedding Models for Supervised Automatic Extraction and Classification of Named Entities in Scientific Acknowledgements}


\author*[1]{\fnm{Nina} \sur{Smirnova}}\email{nina.smirnova@gesis.org}

\author[1]{\fnm{Philipp} \sur{Mayr}}\email{philipp.mayr@gesis.org}


\affil[1]{\orgname{GESIS –- Leibniz Institute for the Social Sciences}, \orgaddress{\street{Unter Sachsenhausen 6-8}, \city{Cologne}, \postcode{50667}, \country{Germany}}}

%\affil[2]{\orgdiv{Department}, \orgname{Organization}, \orgaddress{\street{Street}, \city{City}, \postcode{10587}, \state{State}, \country{Country}}}

%\affil[3]{\orgdiv{Department}, \orgname{Organization}, \orgaddress{\street{Street}, \city{City}, \postcode{610101}, \state{State}, \country{Country}}}

%letter to the reviewer: https://docs.google.com/document/d/1zsyr6gxP6hZ7Femv5wqDZYY9IB4cxMDW/edit


\abstract{Acknowledgments in scientific papers may give an insight into aspects of the scientific community, such as reward systems, collaboration patterns, and hidden research trends. The aim of the paper is to evaluate the performance of different embedding models for the task of automatic extraction and classification of acknowledged entities from the acknowledgment text in scientific papers. We trained and implemented a named entity recognition (NER) task using the Flair NLP framework. The training was conducted using three default Flair NER models with four differently-sized corpora and different versions of the Flair NLP framework. The Flair Embeddings model trained on the medium corpus with the latest FLAIR version showed the best accuracy of 0.79. Expanding the size of a training corpus from very small to medium size massively increased the accuracy of all training algorithms, but further expansion of the training corpus did not bring further improvement. Moreover, the performance of the model slightly deteriorated. Our model is able to recognize six entity types: funding agency, grant number, individuals, university, corporation, and miscellaneous. The model works more precisely for some entity types than for others; thus, individuals and grant numbers showed a very good F1-Score over 0.9. Most of the previous works on acknowledgment analysis were limited by the manual evaluation of data and therefore by the amount of processed data. This model can be applied for the comprehensive analysis of acknowledgment texts and may potentially make a great contribution to the field of automated acknowledgment analysis.
}

\keywords{Natural Language Processing, Names Entity Recognition, Web of Science, Acknowledgement, Text Mining, Flair NLP-Framework}



\maketitle

\section{Introduction}\label{sec:intro}
 
Acknowledgments in scientific papers are short texts where the author(s) \textit{“identify those who made special intellectual or technical contribution to a study that are not sufficient to qualify them for authorship”} \citep[p.~1511]{kassirer_authorship_1991}. \cite{cronin_praxis_1995} ascribe an acknowledgment alongside authorship and citedness to measures of a researcher's scholarly performance: a feature that reflects the researcher’s productivity and impact. 
\cite{giles_who_2004} argue that acknowledgments to individuals, in the same way as citations, may be used as a metric to measure an individual’s intellectual contribution to scientific work. Acknowledgments of financial support are interesting in terms of evaluating the influence of funding agencies on academic research. Acknowledgments of technical and instrumental support may reveal \textit{“indirect contributions of research laboratories and universities to research activities”} \cite[p.~17599]{giles_who_2004}. 

The analysis of acknowledgments is particularly interesting as acknowledgments may give an insight into aspects of the scientific community, such as reward systems \citep{Kazienko2022}, collaboration patterns, and hidden research trends \citep{giles_who_2004, diaz-faes_making_2017}. From the linguistic point of view, acknowledgments are unstructured text data, which through automatic analysis poses research and methodological problems like data cleaning, choosing the proper tokenization method, and whether and how word embeddings may enhance their automatic analysis.

To our knowledge, previous works on automatic acknowledgment analysis were mostly concerned with the extraction of funding organizations and grant numbers \citep{alexandera_this_2021,kayal-etal-2017-tagging, silvello_extracting_2022} or classification of acknowledgment texts \citep{song_kang_timakum_zhang, hubbard_analysis_2022}. 
Furthermore, large bibliographic databases such as Web of Science (WoS)\footnote{\url{http://wokinfo.com/products_tools/multidisciplinary/webofscience/fundingsearch/}} and Scopus selectively index only funding information, i.e., names of funding organizations and grant identification numbers. Consequently, we want to extend that to other types of acknowledged entities: individuals, universities, corporations, and other miscellaneous information. Analysis of the acknowledged individuals provides insight into informal scientific collaboration \citep{georg_rose_2015,kusumegi_dataset_2022}. Acknowledged universities and corporations reveal interactions and knowledge exchange between industry and universities \citep{chen_network_2022}. Entities from the miscellaneous category include other information like project names, which could uncover international scientific collaborations. 

The state-of-the-art named entity recognition (NER) models showed a great performance on the CoNLL-2003 dataset \citep{akbik_contextual_2018, devlin_bert_2018, yamada_luke_2020, yu_named_2020}. CoNLL-2003 corpus \citep{tjong-kim-sang-de-meulder-2003-introduction} is a benchmark dataset for language-independent named entity recognition, i.e., designed to train and evaluate NER models. English data for the corpus were taken from the Reuters corpus. The dataset comprises four types of named entities: person, location, organisation, and miscellaneous. 
However, specific domains require specifically labelled training data. The development of a training dataset for the specific domain is an expensive and time-consuming process since NER usually requires a quite large training corpus.
Therefore, the objective of this paper is to evaluate the performance of existing embedding models for the task of automatic extraction and classification of acknowledged entities from the acknowledgment text in scientific papers using small training datasets or without training data (zero-short approach). 

The present paper is an extended version of the article \citep{smirnova_evaluation_2022}\footnote{In this paper we conducted an additional experiment (Experiment 3) with 2 new corpora (corpus Nos. 3 and 4).} presented at the 3rd Workshop on Extraction and Evaluation of Knowledge Entities from Scientific Documents (EEKE2022)\footnote{\url{https://eeke-workshop.github.io/2022/}}. 
Flair, an open-source natural language processing (NLP) framework \citep{akbik_flair_2019} is used in our study to create a tool for the extraction of acknowledged entities because this library is easily customizable. It offers the possibility of creating a customized Named Entity Recognition (NER) tagger, which can be used for processing and analyzing acknowledgment texts. Furthermore, Flair has shown better accuracy for NER tasks using pre-trained datasets in comparison with many other open source NLP tools\footnote{\url{https://github.com/flairNLP/flair}}. 

In the first experiment (Section~\ref{sec:def_train}) we trained and implemented a NER task using three default Flair NER models with two differently-sized corpora\footnote{The release 0.9 (\url{https://github.com/flairNLP/flair/releases/tag/v0.9}) was used in the experiments 1 and 2. Experiment 3 was performed using release 0.11 (\url{https://github.com/flairNLP/flair/releases/tag/v0.11})}. All the descriptions of the Flair framework features refer to the releases 0.9 and 0.11. The models were trained to recognize six types of acknowledged entities: funding agency, grant number, individuals, university, corporation, and miscellaneous. The model with the best accuracy can be applied for the comprehensive analysis of the acknowledgment texts.
In Experiments 2 and 3 we performed additional training with altered training parameters or altered training corpora (Sections~\ref{sec:add_train} and \ref{sec:exp_3}).
Most of the previous works on acknowledgment analysis were limited by the manual evaluation of data and therefore by the amount of processed data \citep{giles_who_2004, paul_hus_all_2017,paul_hus_des_2019, Mccain2017}. Furthermore, \cite{thomer_weber_2014} argues that using named entities can benefit the process of manual document classification and evaluation of the data. Therefore, a model that is capable of extracting and classification of different types of entities may potentially make a significant contribution to the field of automated acknowledgment analysis.  

\subsection*{Research questions}
In this paper, we address the following research questions:
\begin{itemize}
    \item \textbf{RQ1:} Is the few-shot or zero-shot approach able to identify predefined acknowledged entity classes? 
    \item \textbf{RQ2:} Which of the Flair default NER models is more suitable for the defined task of extraction and classification of acknowledged entities from scientific acknowledgments using a small training dataset?
    \item \textbf{RQ3:} How does the size of the training corpus affect the training accuracy for different NER models? 
\end{itemize}

Creating a training dataset for supervised learning is a time-consuming and expensive task, since as a rule, such a model requires a reasonably large amount of training data. Annotation is a crucial moment, as wrongly annotated data will deteriorate training results. Therefore, more than one annotator is usually required to provide credible results. That is why it is of interest to test if the existing NER models can provide reasonable accuracy while using small or no training data.

\section{Background and Related work}\label{sec:background}

Research in the field of acknowledgments analysis has been carried out since the 1970s. The first typology of acknowledgments was proposed by \cite{mackintosh_1972} \citep[as cited in][]{Cronin1995TheSC} and comprised three categories:  facilities, access to data, and help of individuals. \cite{mccain_1991} distinguished five types of acknowledgements: research-related information, secondary access to research-related information, specific research-related communication, general peer communication, and technical or clerical support. \cite{cronin_praxis_1995} defined three broad categories: resource-, procedure- and concept-related. \cite{mejia_using_2018} developed a four-level classification based on sponsored research field: change maker, incremental, breakthrough, and matured.  

\cite{doehne_how_2023} distinguished acknowledgements from the perspective of appreciation of influential scholars and defined two axes: scientific influence and institutional influence. Scientific influence refers to the productiveness and creativity of the researcher, while institutional influence is associated with the scholar's administrative position in the scientific community.  

\cite{wang_funding_2011} investigated the connection between research funding and the development of science and technology using acknowledgments from articles from the field of nanotechnology.
\cite{georg_rose_2015} studied informal cooperation in academic research. The analysis revealed generational and gender differences in informal collaboration. The authors claim that information from informal collaboration networks makes better predictions of the academic impact of researchers and articles than information from co-author networks.
\cite{mejia_using_2018} argued that the classification of funders could be useful in developing funding strategies for policymakers and funders.

\cite{doehne_how_2023} manually investigated acknowledgement sections of papers, which were published or preprinted in association with the Cowles Foundation between early 1940 and 1970 to trace the influence of the informal social structure and academic leaders on the early acceptance of scientific innovations. Blockmodelling was applied to the acknowledgement data. Their analysis showed that the adoption of scientific innovations was partly influenced by the social structure and by the scientific leaders at Cowles.

\subsection{Recent advances in NER}\label{subsec:rec_ner}

Named Entity Recognition (NER) is a form of NLP that aims to extract named entities from unstructured text and classify them into predefined categories. A named entity is a real-world object that is important for understanding the text.  Current approaches in NER can be distinguished into supervised and unsupervised tasks. In a supervised NER a model is trained using a labelled dataset.  
This training dataset or corpus is usually split into several datasets: training set, test set, and validation set. NER models require corpora with semantic annotation, i.e., metadata about concepts attached to unstructured text data. The annotation process is crucial as insufficient or redundant metadata can slow down and bias a learning process \citep[Chapter~1]{pustejovsky_natural_2012}. 

Supervised NER mainly relays on machine learning or deep learning  methods. The state-of-the-art models are based on deep recurrent models, convolution-based, or pre-trained transformer architectures \citep{iovine_cyclener_2022}.
Thus, \cite{akbik_contextual_2018} proposed a new character-based contextual string embeddings method. This approach passes a sequence of characters through the character-level language model to generate word-level embeddings. The model was pre-trained on large unlabeled corpora. The training was carried out using a character-based neural language model together with a Bidirectional LSTM (BiLSTM) sequence-labelling model. This approach generates different embeddings for the same word depending on its context and showed good results on downstream tasks such as NER.
\cite{devlin_bert_2018} presented BERT (Bidirectional Encoder Representations Transformers), a transformer-based language representation model that models the representation of contextualized word embeddings. BERT showed superior results on downstream tasks using different benchmarking datasets.  Later, \cite{liu_roberta_2019} performed an optimization of the BERT model and introduced RoBERTa (Robustly Optimized BERT Pretraining Approach). RoBERTa was evaluated on three benchmarks and demonstrated massive improvements over the reported BERT performance. 

Currently, several domain-specific models have been developed. Thus, \cite{beltagy_scibert_2019} released SciBERT a BERT-based language model pre-trained on a large number of unlabeled scientific articles from the computer science and biomedical domains. SciBERT showed improvements over BERT on several downstream NLP tasks, including NER. Recently, \cite{shen_sscibert_2022} introduced the SsciBERT, a language model based on BERT and pre-trained on abstracts published in the Social Science Citation Index (SSCI) journals. The model showed good results in discipline classification and abstract structure-function recognition in articles from the social sciences domain.

Unsupervised methods are often based on lexicons or predefined rules. Thus, \cite{etzioni_unsupervised_2005} uses lists of patterns and domain-specific rules to extract named entities. \cite{eftimov_rule-based_2017} developed a rule-based NER model to extract dietary information from scientific publications. Evaluation of the model performance showed good results. Opposed to previous unsupervised NER approaches, \cite{iovine_cyclener_2022} proposed a cycle-consistency approach for NER (CycleNER). CycleNER is unsupervised and does not require parallel training data. The method showed 73\% of supervised performance on CoNLL03.

\begin{table}[ht]
\centering
\begin{tabular}{p{0.12\linewidth} | p{0.22\linewidth} | p{0.15\linewidth} | p{0.2\linewidth} |p{0.14\linewidth}}
\toprule
                       Paper &                                                 Area of application and aim of the study &                              Corpus &                                                                Entities &                                           Methods and tools \\
\midrule
   Giles and Councill (2004) &                               extraction of acknowledged entities form acknowledgements &                            CiteSeer &      Funding agencies, Companies, Educational Institutions, Individuals & SVM for extracting entities and their manual classification \\
     Thomer and Weber (2014) &                                  using NER to improve classification of acknowledgements &         PubMed Central’s Open Access &                    persons, locations, organizations, and miscellaneous &                          4-class Stanford Entity Recognizer \\
         Kayal et al. (2017) &                                  extraction of funding information from acknowledgements &         PubMed Central’s Open Access &                                                  funding bodies, grants &                                            CRF, HMM, MaxEnt \\
           Kenekayoro (2018) &                             extraction of  biography information from academic biographies &                               ORCID & Award, Location, Organization, Person, Position, Specialization, Others &                                                         SVM \\
Alexandera and Vries (2021)  &                                  extraction of funding information from acknowledgements & TU Delft’s institutional repository &                                                  funding bodies, grants &              SpaCy dependency parser +  regular expressions \\
        Jiang et al. (2022)  & extraction of scientific software from scientific articles (full texts) in bioinformatics &             bioinformatics journals &                                                                      &                                            EnsembleSVMs-CRF \\
         Borst et al. (2022) &                                  extraction of funding information from acknowledgements &                            EconStor &                                                  funding bodies, grants &                                                    Haystack \\
   Kusumegi and Sano (2022)  &               extraction and linking of acknowledged individuals from acknowledgements  &                                PLOS &                                                             individuals &                           Stanford CoreNLP NER tagger + MAG \\
\bottomrule
\end{tabular}
\caption{Overview of works on NER in scientometrics}
\label{tab:ner_overview}
\end{table}

\subsection{NER in scientometrics analysis}\label{subsec:scim_ner}

Named entities are widely used in scientometrics analysis. Thus, \cite{kenekayoro_identifying_2018} developed a supervised method for the automatic extraction of named entities from academic bibliographies. The aim of the study was to create a database containing unified academic information about individuals to help in expert finding. A labeled training dataset was developed using biographies extracted from ORCID\footnote{\url{https://orcid.org/}}. The authors tested several models for NER. The Support Vector Machine classification algorithm (SVM) showed the best performance.

\cite{jiang_refinement_2022} proposed a strategy for the identification of software in scientific bioinformatics publications using the combination of SVM and CRF (Conditional Random Field). Application of the method to the sample of articles from bioinformatics domains allowed them to observe interesting patterns in using software in scientific research.

\cite{kusumegi_dataset_2022} analysed scholarly relationships by analysing acknowledged individuals from the acknowledgments statements from eight open-access journals. Individuals were extracted using the Stanford CoreNLP NER tagger. In the next steps, scholars were identified among the extracted individuals by mapping them to the Microsoft Academic Graph (MAG).

We are aware of several works on automated information extraction from acknowledgments. \cite{giles_who_2004} developed an automated method for the extraction and analysis of acknowledgment texts using regular expressions and SVM. Computer science research papers from the CiteSeer digital library were used as a data source. Extracted entities were analysed and manually assigned to the following four categories: funding agencies, corporations, universities, and individuals.

\cite{thomer_weber_2014} used the 4-class Stanford Entity Recognizer \citep{finkel-etal-2005-incorporating} to extract persons, locations, organizations, and miscellaneous entities from the collection of bioinformatics texts from PubMed Central's Open Access corpus. The aim of the study was to determine an approach to \textit{"increase the speed of ... classification without sacrificing accuracy, nor reliability"} \citep[p.~1134]{thomer_weber_2014}. 

\cite{kayal-etal-2017-tagging} introduced a method for extraction of funding organizations and grants from acknowledgment texts using a combination of sequential learning models: conditional random fields (CRF), hidden markov models (HMM), and maximum entropy models (MaxEnt). The final model contained pooled outputs from the models used. 

\cite{alexandera_this_2021} proposed AckNER, a tool for extracting financial information from the funding or acknowledgment section of a research article. AckNER works with the use of dependency parse trees and regular expressions and is able to extract names of the organisations, projects, programs, and funds, as also numbers of contracts and grants \footnote{AckNER showed better performance as Flair, but is specifically designed to recognize two types of acknowledged entities \citep{alexandera_this_2021}, which was insufficient for the present project.}. 

Following, \cite{silvello_extracting_2022} applied a question-answering (QA) based approach to identify funding information in acknowledgments texts. This approach performs similarly to AckNER and requires a smaller set of training and test data. 

Table~\ref{tab:ner_overview} shows an overview of works on NER in scientometrics. Overall, previous works on the extraction of named entities from acknowledgements texts were mostly concerned with the extraction of funding information, i.e., only names of funding bodies and grant numbers, or extraction and linking of individuals. The special issue by \cite{Chengzhi23} provided a recent overview of current works in the extraction of knowledge entities.

To the best of our knowledge the work of \cite{giles_who_2004} is the only attempt to extract and categorise multiple acknowledged entities. Nevertheless, entities were extracted using the SVM algorithm but the classification of entities themselves was produced manually, which limited the number of acknowledgement texts to be analysed. Furthermore, as far as we know, there was no research done concerning the evaluation of embedding models for extraction of information from acknowledgement texts and no tool for automatic extraction of different kinds of acknowledged entities was developed.

\section{Method}\label{sec:usecase}

\begin{figure}[h!]
\centering
  \includegraphics[width=0.8\textwidth]{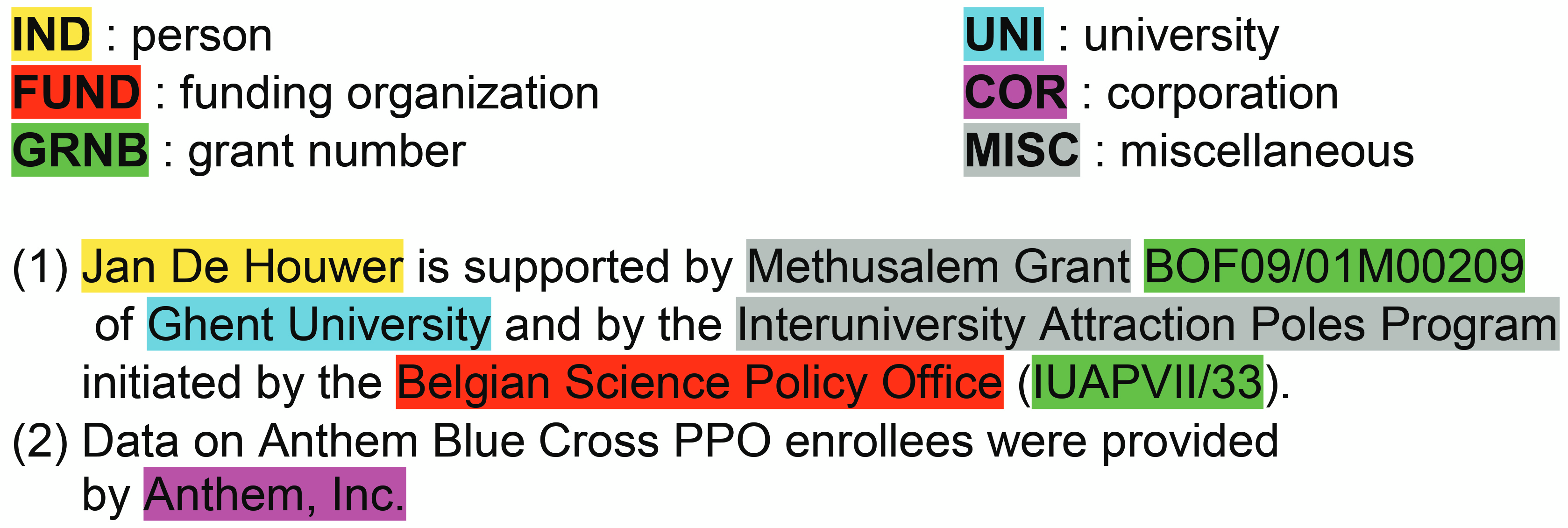}
  \caption{An example of acknowledged entities. Each entity type is marked with a distinct color.}
  \label{fig:example_ackn}
  %\vspace{-4mm}
\end{figure}

In the present paper, different models for extraction and classification of acknowledged entities supported by the Flair NLP Framework were evaluated. The choice of classification was inspired by \posscite{giles_who_2004} classification: funding agencies (FUND), corporations (COR), universities (UNI), and individuals (IND). For our project, this classification was enhanced with the miscellaneous (MISC) and grant numbers (GRNB) categories. The GRNB category was adopted from WoS funding information indexing. The entities in the miscellaneous category could provide useful information, but cannot be ascribed to other categories, e.g., names of projects and names of conferences. Figure~\ref{fig:example_ackn} demonstrates an example of acknowledged entities of different types. To the best of our knowledge, Giles and Councill's classification is the only existing classification of acknowledged entities and therefore can be applied to the NER task. Other works on acknowledgment analysis focused on the classification of acknowledgment texts. 

\subsection{The Flair NLP Framework}\label{subsec:flair}

Flair is an open-sourced NLP framework built on PyTorch \citep{paszke_pytorch_2019}, which is an open-source machine learning library. \textit{“The core idea of the framework is to present a simple, unified interface for conceptually very different types of word and document embeddings”} \citep[p.~54]{akbik_flair_2019}. Flair has three default training algorithms for NER which were used for the first experiment in the present research: a) NER Model with Flair Embeddings (later on Flair Embeddings) \citep{akbik_contextual_2018}, b) NER Model with Transformers (later on Transformers) \citep{schweter_flert_2020}, and c) Zero-shot NER with TARS (later on TARS) \citep{halder_task-aware_2020}~\footnote{New transformer models as SciBERT or SsciBERT were not evaluated in this study, as the objective of the study is to evaluate the performance of the Flair default models.}.

The Flair Embeddings model uses stacked embeddings, i.e., a combination of contextual string embeddings \citep{akbik_contextual_2018} with a static embeddings model. This approach will generate different embeddings for the same word depending on its context. Stacked embedding is an important Flair feature, as a combination of different embeddings might bring better results than their separate uses \citep{akbik_flair_2019}. 

The Transformers model or FLERT-extension (document-level features for NER) is a set of settings to perform a NER on the document level using fine-tuning and feature-based LSTM-CRF with the multilingual XML-RoBERTa transformer model \citep{schweter_flert_2020}. 

The TARS (task-aware representation of sentences) is a transformer-based model, which allows performing training without any training data (zero-shot learning) or with a small dataset (few-short learning) \citep{halder_task-aware_2020}. The TARS approach differs from the traditional transfer learning approach in the way that the TARS model also considers semantic information captured in the class labels themselves. For example, for analyzing acknowledgments, class labels like \textit{funding organization} or \textit{university} already carry semantic information. 
 %\vspace{-2mm}
 
\subsection{Training Data}
The Web of Science (WoS) database was used to harvest the training data (funding acknowledgments)\footnote{The present research was conducted in scopes of two projects: MinAck (\url{https://kalawinka.github.io/minack/}) and SEASON (\url{https://github.com/kalawinka/season}). Corpora Nos.1, 2, and 3 were created for the MinAck project and serve the purpose of a general evaluation of the impact of the size of the training corpus on the model performance. Corpus No.4 was designed specifically for the SEASON project in the hope of improving the recognition of Indian funding information. The project SEASON aims to get insight into German-Indian scientific collaboration. Our other corpora mainly contain papers published by European institutions. That is why we enhance Corpus 4 with the papers published by Indian institutions. }. From 2008 on, WoS started indexing information about funders and grants. WoS uses information from different funding reporting systems such as researchfish\footnote{\url{https://mrc.ukri.org/funding/guidance-for-mrc-award-holders/researchfish/}}, Medline\footnote{\url{https://www.nlm.nih.gov/bsd/funding_support.html}} and others. 
As WoS contains millions of metadata records \citep{singh2021}, the data chosen for the present study was restricted by year and scientific domain (for the corpora Nos. 1, 2, and 3) or additionally by the affiliation country (for corpus No.4). To construct corpora Nos. 1-3 records from four different scientific domains published from 2014 to 2019 were considered: two domains from the social sciences (sociology and economics) and oceanography and computer science. Different scientific domains were chosen since previous work on acknowledgment analysis revealed the relations between the scientific domain and the types of acknowledged entities, i.e., acknowledged individuals are more characteristic of theoretical and social-oriented domains. At the same time, information on technical and instrumental support is more common for the natural and life sciences domains \citep{diaz-faes_making_2017}. Only the WoS record types \textit{“article”} and \textit{“review”} published in a scientific journal in English were selected; then 1000 distinct acknowledgments texts were randomly gathered from this sample for the training dataset. Further different amounts of sentences containing acknowledged entities were distributed into the differently-sized training corpora. Table~\ref{tab:ackn_corpus_count} demonstrates the number of sentences in each set in the four corpora. We selected only sentences that contain an acknowledged entity, regardless of the scientific domain. Table~\ref{tab:train_corpus_disc} contains the number of sentences and texts from each scientific domain in the training corpora\footnote{Corpus No.4 is not in Table 2, because the corpus contains additional acknowledgment texts from articles with Indian affiliations regardless of the scientific domain and therefore contains different scientific domains.}. The same article can belong to several scientific domains, therefore, the number of sentences and texts in Tables~\ref{tab:ackn_corpus_count} and \ref{tab:train_corpus_disc} does not match. Corpus No.4 was designed in such a way that all the training data from the Corpus No.3 was enhanced with acknowledgments texts from the articles that have Indian affiliations regardless of scientific domain or publication date.

\begin{table}[h!]
%\begin{tabular}{ |p{1,2cm}||p{1,5cm}|p{1,5cm}|p{1,5cm}|p{1cm}| }
\begin{tabular}{ lllll}
 \hline
 Corpus No.    & Training set (train)    & Test set (test) &   Validation set (dev) & \textbf{Total}\\
 \hline
 1 &   29 / 27  & 10 / 10   & 10 / 10 & \textbf{49 / 47}\\
 2 & 339 / 282 & 165 / 150 &  150 / 136 & \textbf{654 / 441}\\
 3 & 784 / 657 & 165 / 150 &  150 / 136 & \textbf{1099 / 816}\\
 4 & 1148 / 885 & 165 / 150 &  150 / 136 & \textbf{1463 / 1044}\\
 \hline
\end{tabular}
\caption{\label{tab:ackn_corpus_count} Number of sentences / texts in the training corpora.}
%\vspace{-8mm}
\end{table}

\begin{table}[h!]
\begin{tabular}{lllll}
 \hline
 Corpus No. & Oceanography & Economics & Social Sciences & Computer Science\\
 \hline
 1 &   13 / 13  & 3 / 3   & 20 / 20 & 16 / 14\\
 2 & 127 / 75 & 92 / 58 &  351 / 234 & 173 / 129\\
 3 & 175 / 112 & 128 / 89 & 590 / 434  & 333 / 269\\
 \hline
\end{tabular}
\caption{\label{tab:train_corpus_disc} Number of sentences / texts from each scientific domain in the training corpora.}
%\vspace{-6mm}
\end{table}

\begin{figure}[h!]
\centering
  \includegraphics[width=1\textwidth]{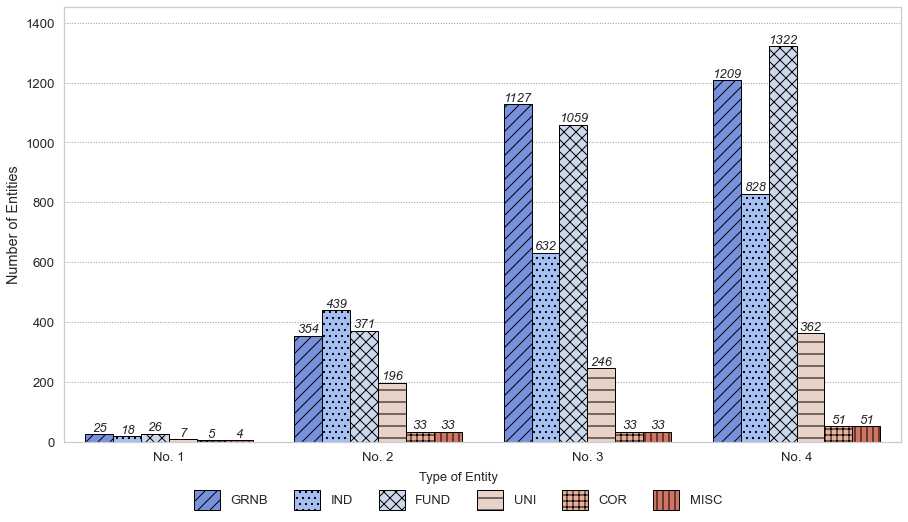}
  \caption{The distribution of acknowledged entities  in the training corpora.}
  \label{fig:corpora_counts}
  %\vspace{-3mm}
\end{figure}

\begin{figure}[h!]
\centering
  \includegraphics[width=1\textwidth]{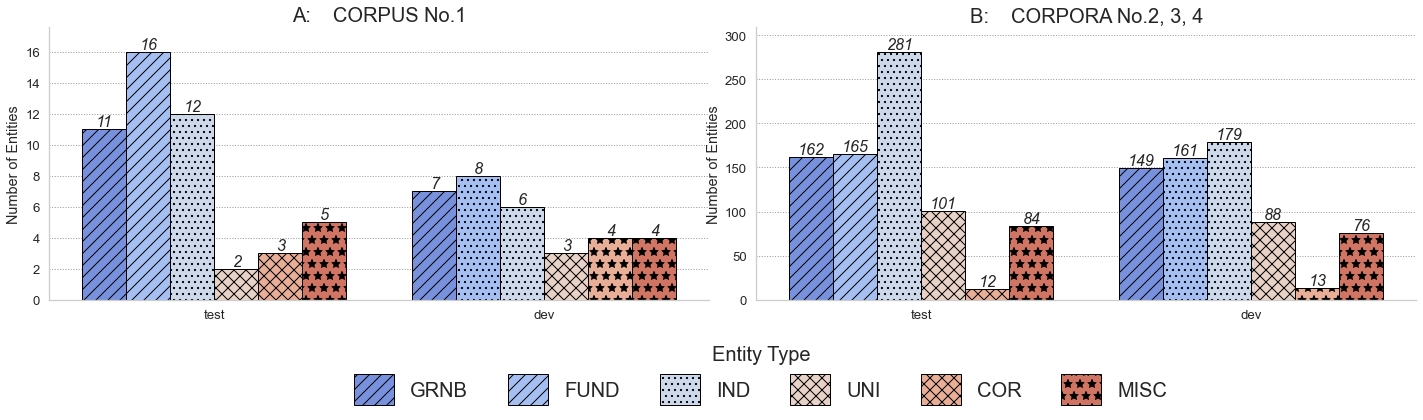}
  \caption{The distribution of acknowledged entities in the test and validation corpora.}
  \label{fig:test_dev_ent_freq}
  %\vspace{-3mm}
\end{figure}

Preliminary analysis of the WoS data showed that the indexing of WoS funding information has several issues. The WoS includes only acknowledgments containing funding information; therefore, not every WoS entry has an acknowledgment, individuals are not included, and indexed funding organizations are not divided into different entity types like universities, corporations, etc. Therefore, the existing indexing of funding organizations is incomplete. Furthermore, there is a disproportion between the occurrences of acknowledged entities of different types. Thus, the most frequent entity types in the dataset with the training data are IND, FUND and GRNB, followed by UNI and MISC. COR is the category most underrepresented in the data set. Consequently, there are different amounts of entities of different types in the training corpora (as Figure~\ref{fig:corpora_counts} demonstrates), which might have influenced the training results. Training with the corpora Nos. 2, 3, and 4 was evaluated on the same training and validation datasets to ensure plausible accuracy (Figure~\ref{fig:test_dev_ent_freq}-B). However, training with corpus No.1 was evaluated with the smaller test and validation sets, as corpus No.1 contains a smaller number of sentences (Figure~\ref{fig:test_dev_ent_freq}-A).

\subsection{Data Annotation}

\begin{figure}[h!]
\centering
  \includegraphics[width=0.7\textwidth]{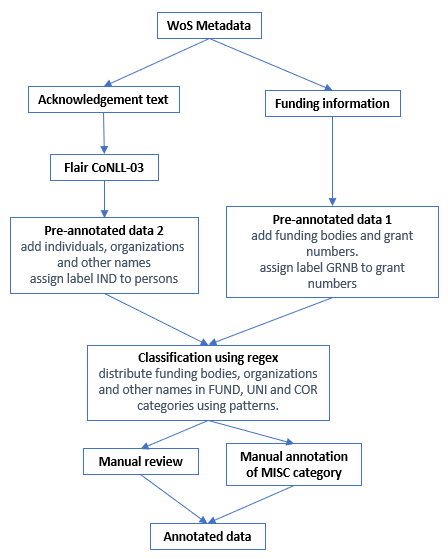}
  \caption{Annotation flowchart.}
  \label{fig:annotation}
  %\vspace{-3mm}
\end{figure}

The training corpus was annotated with six types of entities. As WoS already contains some indexed funding information, it was decided to develop a semi-automated approach for data annotation (as Figure~\ref{fig:annotation} demonstrates) and use indexed information provided by WoS, therefore, grant numbers were adopted from the WoS indexing unaltered. 

Flair has a pre-trained 4-class NER Flair model (CoNLL-03)\footnote{\url{https://github.com/flairNLP/flair}}. The model can predict four tags: PER (person name), LOC (location), ORG (organization name), and MISC (other names). As Flair showed adequate results in the extraction of names of individuals, it was decided to apply the pre-trained 4-class CoNLL-03 Flair model to the training dataset. Entities that fell into the PER category were added as the IND annotation to the training corpus. Furthermore, we noticed that some funding information was partially correctly extracted into the ORG and MISC categories. Therefore, WoS funding organization indexing and entities from the ORG and MISC categories were adopted and distinguished between three categories (FUND, COR, and UNI) using regular expressions. In addition, the automatic classification of entities was manually examined and reviewed. Mismatched categories, partially extracted entities, and not extracted entities were corrected. Acknowledged entities, which fall into the MISC category, were manually annotated by one annotator. In the miscellaneous category entities referring to names of the conferences and projects were included.

\section{Experiments}\label{sec:exp}

{In the present paper, we evaluated three default Flair NER models with four differently-sized corpora. In total, we performed three experiments. In the first experiment, models with the default parameter were evaluated using corpora Nos. 1 and 2. In the second experiment, we evaluated Flair Embeddings and Transformers model with altered training parameters and corpus No.2. In the third experiment, the first experiment was replicated with corpora Nos. 3 and 4. 
\subsection{Experiment 1} \label{sec:def_train}
 
In the first experiment, we tested the TARS model zero-shot and few-shot scenarios (with corpus No. 1), as well as the performance of two default FLAIR models (Flair Embeddings and Transformers) with corpus No.2. Additionally, the performance of Flair Embeddings and Transformers models was tested with the corpus No.1 
The training was conducted with the recommended parameters for all algorithms, as Flair developers specifically ran various tests to find the best hyperparameters for the default models. For the few-shot TARS, the training was conducted with the small dataset (corpus No.1), and for Transformers and Flair Embeddings with a larger dataset (corpus No.2). 

The Flair Embeddings model was initiated as a combination of static and contextual string embeddings. We applied GloVe \citep{pennington_glove_2014} as a static word-level embedding model.
Thus, in our case, stacked embeddings comprise GloVe embeddings, forward contextual string embeddings, and backward contextual string embeddings. The model was trained with the recommended parameters: the size of mini-batches was set to 32 and the maximum number of epochs was set to 150.

For Transformers, training was initiated with the RoBERTa model \citep{liu_roberta_2019}. For the present paper, a fine-tuning approach was used. The fine-tuning procedure consisted of adding a linear layer to a transformer and retraining the entire network with a small learning rate. We used a standard approach, where only a linear classifier layer was added on the top of the transformer, as adding the additional CRF decoder between the transformer and linear classifier did not increase accuracy compared with this standard approach  \citep{schweter_flert_2020}. The chosen transformer model uses subword tokenization. We used the mean of embeddings of all subtokens and concatenation of all transformer layers to produce embeddings. The context around the sentence was considered. The training was initiated with a small learning rate using the Adam Optimisation Algorithm \citep{adam}.

The TARS model requires labels to be defined in a natural language. Therefore, we transformed our original coded labels into the natural language: FUND - “Funding  Agency”, IND - “Person”, COR - “Corporation”, GRNB - “Grant Number", UNI - “University”, and MISC - “Miscellaneous”. The training for the few-shot approach was initiated with the TARS NER model \citep{halder_task-aware_2020}.

\subsubsection{Results}

Overall, the training demonstrated mixed results.  
Table~\ref{tab:tars_zero_acc} shows training results with corpus No.1 and the TARS zero-shot approach. GRNB showed adequate results by training with Flair Embeddings and TARSfew-shot models. IND was the best-recognized entity by training with Flair Embeddings and TARS (both zero- and few-shot) with an F1-score of 0.8 (Flair Embeddings) and 0.86 (TARS) respectively. Training with Transformers was not successful for IND with an F1-score of 0. In general, transformers were a less efficient algorithm for training with a small dataset with an overall accuracy of 0.35. FUND demonstrated not satisfactory results with F1-score of less than 0.5 for all models. Entity types MISC, UNI, and COR showed the worst results with the F1-score equal to zero for all algorithms. The low accuracy for MISC, UNI, and COR resulted in low overall accuracy for all algorithms. Overall, training with corpus No.1 showed insufficient results for all algorithms. Flair Embeddings and TARS showed better accuracy compared to Transformers.

Figure~\ref{fig:prim_train_comb} shows the training results with corpus No.2. Similar to the training with corpus No.1, IND and GRNB are the best-recognized categories. The best results for IND and GRNB demonstrated Flair embeddings with an F1-score of 0.98 (IND) and 0.96 (GRNB). TARS achieved the best results for FUND with an F1-score of 0.77 against 0.71 for Flair Embeddings and 0.68 for Transformers. Miscellaneous demonstrated the worst accuracy for Flair Embeddings (0.64) and Transformers (0.49), while for TARS the worst accuracy lies in the COR category with an F1-score of 0.54. The best result for UNI showed Flair Embeddings with an F1-score over 0.7. The COR category showed a decent precision of 0.88 with Flair Embeddings but a low recall of 0.58 which resulted in a low F1-Score (0.7)\footnote{Accuracy metrics by type of entity and total accuracy for all experiments can be found in Appendixes~\ref{app:class_accuracy} and \ref{app:accuracy}}. 

\begin{figure}[h!]
\centering
  \includegraphics[width=1\textwidth]{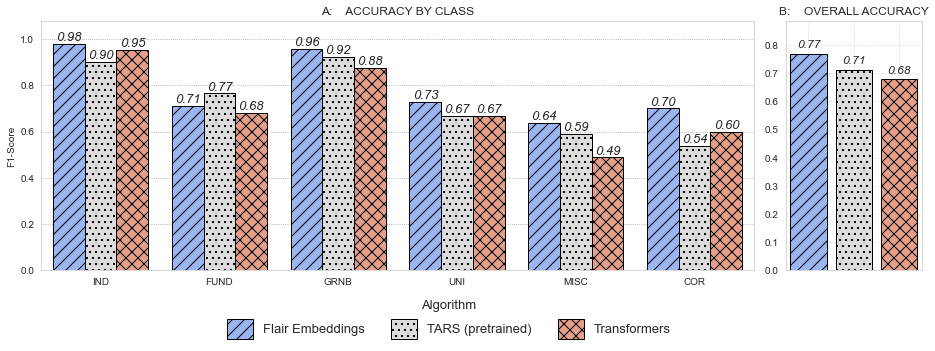}
  \caption{The training results with the training corpus No.2. Figure A comprises diagrams with the  F1-scores of the training with three algorithms for each label class. Figure B depicts the total accuracy of training algorithms.}
  \label{fig:prim_train_comb}
  %\vspace{-4.5mm}
\end{figure}

Training with corpus No.2 showed a significant improvement in training accuracy (Figure~\ref{fig:prim_train_comb}-B). Overall, Flair Embeddings was more accurate than other training algorithms, although training with TARS showed better results for the FUND category. The Transformers showed the worst results during training. 

Additionally, a zero-shot approach was tested for the TARS model on corpus no.1. The model was able to successfully recognize individuals, but struggled with other categories, as Table~\ref{tab:tars_zero_acc} demonstrates. The total accuracy of the model comprises 0.23.

\begin{table}[h!]
\begin{tabular}{ llllllll }
 \hline
 Algorithm & FUND & GRNB & IND &  UNI & COR & MISC & accuracy \\
 \hline
 TARS (zero-shot) & 0.23  & 0.33   & 0.86 & 0 & 0& 0 & 0.23 \\
 TARS (few-shot) & 0.32 & 0.76 & 0.86 & 0 & 0& 0 & 0.35 \\
 Flair Embeddings & 0.42 & 0.61 & 0.80 & 0 & 0& 0 &  0.35 \\
 Transformers & 0.30 & 0.40  & 0 & 0& 0 & 0 & 0.15 \\
 \hline
\end{tabular}
\caption{\label{tab:tars_zero_acc}  F1-scores of the training with three algorithms for each label class with Corpus No. 1.}
\end{table}

\subsection{Experiment 2} \label{sec:add_train}

Our first hypothesis to explain the pure model performance for the FUND, COR, MISC, and UNI categories is their semantic proximity that prevents successful recognition. Entities of these categories are often used in the same context. %Training with another dataset, which combines these categories into one category might improve the results. 
To examine this hypothesis, we conducted an experiment using Flair Embeddings with the dataset containing three types of entities: IND, GRNB, and ORG. The MISC category was excluded from the training, as one of the aims of the present research is to extract information about acknowledged entities, and the MISC category contains only additional information. The new ORG category was established, which includes a combination of entities from the FUND, COR, and UNI categories. The training was performed with exactly the same parameters as training with the Flair Embeddings model in Experiment 1 (Section~\ref{sec:def_train}). 

The UNI and COR categories, though, have distinct patterns. 
In this case, the low performance of the models for the COR and UNI categories could be explained by the small size of the training sample that contains these categories (see Figure~\ref{fig:corpora_counts}). Thus, the model was not able to identify patterns because of the lack of data.

Secondly, low results for FUND, COR, MISC, and UNI categories might also lie in the nature of the miscellaneous category, as some entities that fall into this category are semantically very close to the FUND and COR categories. As a result, training without a MISC category might potentially show better performance.
To examine this hypothesis, we conducted training with Flair Embeddings with a dataset excluding the MISC category, i.e., with five entity types. Training results are shown in Figure~\ref{fig:add_train_comb}-A. 

Additionally, the problem might lie in the nature of the training algorithms that were used. On the one hand, Flair developers claimed Transformers to be the most efficient algorithm \citep{schweter_flert_2020}. On the other, the stacked embeddings are an important feature of the Flair tool, as a combination of different embeddings might bring better results than their separate uses \citep{akbik_flair_2019}. Thus, the combination of the Transformer embeddings model with the contextual string embeddings might improve the model performance.
Thus, for the third additional training, we combined contextual string embeddings with FLERT parameters. 

\subsubsection{Results}
Results of the training are represented in Figure~\ref{fig:add_train_comb}. During the training with three types of entities (Figure~\ref{fig:add_train_comb}-B) IND and GRNB still achieved high F1-scores of 0.96 (IND) and 0.95 (GRNB). Nevertheless, ORG gained only an F1-score of 0.64, which is worse than the previous results with six entity types. 
The results of the training with five types of entities were quite similar to those achieved during the training with six types of entities. FUND and UNI categories showed a small improvement in precision, recall, and F1 score compared to training with 6 types of entities with Flair Embeddings. At the same time, the performance of the COR category deteriorated noticeably (0.6 vs. the previous 0.7).
The improvement in overall accuracy (Figure~\ref{fig:add_train_comb}-D) (0.80 vs. the previous 0.77) could be explained by the fact that the MISC category was not present in this training and could not affect overall accuracy with its low F1-score. 

As Figure~\ref{fig:add_train_comb}-C demonstrates, training with Flair Embeddings and RoBERTa showed no improvements compared to the results of the primary training with Transformers and worse performance compared with Flair Embeddings. As in Experiment 1, the COR category achieved high precision but low recall, resulting in a low F1-score (0.67). For some categories (COR and GRNB) Flair Embeddings combined with RoBERTa performed better than Transformers but still worse than Flair Embeddings. 

\begin{figure}[h!]
\centering
  \includegraphics[width=0.8\textwidth]{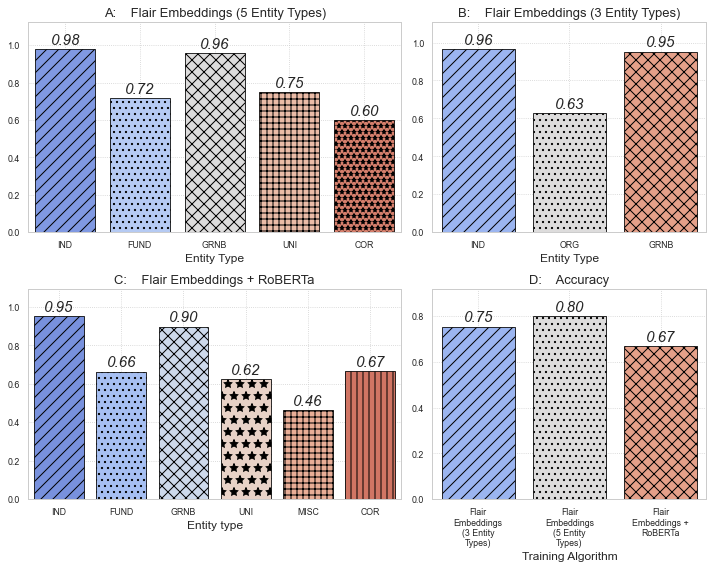}
  \caption{The results of Experiment 2. Figures A, B, and C comprise diagrams with the F1-scores of the training with three algorithms for each label class. Figure D represents the total accuracy of the training algorithms.}
  \label{fig:add_train_comb}
\end{figure}

\subsection{Experiment 3} \label{sec:exp_3}
The results of experiment 2 showed that altering the training parameters and decreasing the number of entity classes does not improve the model accuracy. We assume that increasing the size of the training corpus would improve the performance of entities with low recognition accuracy. Therefore, for this experiment, we designed two corpora with an increased number of acknowledged entities.

As the Flair Embeddings algorithm trained with Corpus No.2 showed the best performance, it was of interest if the increased training data will outperform its accuracy score.  Training in Experiments 1 and 2 was carried out using Flair version 0.9. As Flair recently updated to version 0.11, we used this newest version for the following training. The training was carried out with exactly the same parameters as the training with the Flair Embeddings model in Experiment 1 (Section 3.1).
To achieve comparable results we also retrained, for now, the best model (Flair Embeddings with Corpus No.2) with the Flair 0.11. 

\subsubsection{Results}
Results of the training are represented in Figure~\ref{fig:exp3_res}. Retraining of the original model with the Flair 0.11 Figure~\ref{fig:exp3_res}-B showed slightly better performance (0.79 vs. 0.77) than training with version 0.9. In general, no huge differences in accuracy were found during training with extended corpora.

\begin{figure}[h!]
\centering
  \includegraphics[width=1\textwidth]{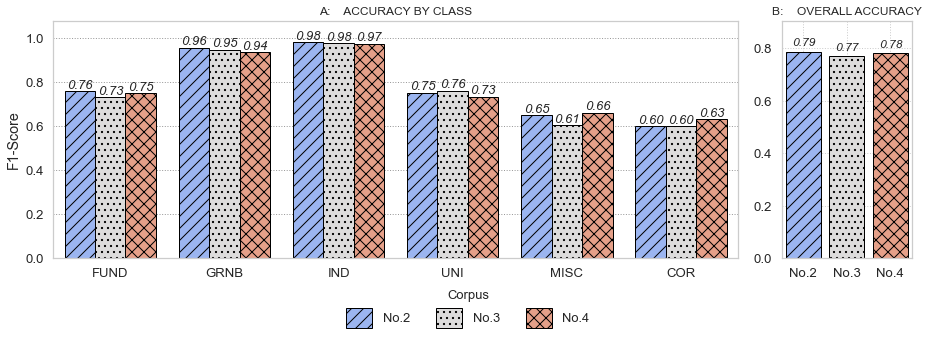}
  \caption{The results of Experiment 3. Figure A comprises diagrams with the F1-scores of training with three corpora for each label class. Figure B represents the total accuracy of the training.}
  \label{fig:exp3_res}
  %\vspace{-4.5mm}
\end{figure}

Overall, the best F1-Score for the FUND category (0.77) was reached with the TARS algorithm and corpus No.2. COR gained the best accuracy (0.7) with Flair Embeddings and corpus No.2 using Flair version 0.9. The GRNB category showed the best performance (0.96) with Flair Embeddings trained on the corpus with five types of entities (Flair Embeddings 5 Ent). The best F1-Score of the IND category was achieved with Flair Embeddings trained on corpus No.2 with Flair version 0.11. MISC performed the best (0.66) with Flair Embeddings trained on Corpus No.4 with Flair version 0.11. The best accuracy of the UNI category was achieved with Flair Embeddings trained on corpus No.3 with Flair version 0.11. In general, the best overall accuracy of 0.79 (for six entity types) had the Flair Embeddings model trained on corpus No.2 with Flair version 0.11.

\section{Discussion}\label{disc_add_train}

As expected, Experiment 1 showed a large improvement in accuracy for all algorithms when the size of a training corpus was increased from 49 to 654 sentences. However, further enlargement of the corpus (in Experiment 3) did not make any progress. Some types of entity, such as IND and GRNB, showed great performance (GRNB with an F1-Score of 0.96 or IND with 0.98) with the small training samples, i.e., 354 entities from the GRNB category or 439 entities from the IND category. At the same time, training with a sample of 1322 labelled funding organisations achieved an F1-Score of only 0.75.

The TARS model is designed to perform NER with small or no training data. In experiment 1, TARS without training data was able to extract individuals with quite high accuracy (F-1 score of 0.86). TARS trained with the small corpus (No. 1) did not show improvement in the F-1 score of individuals, but greatly improved the F-1 score of the GRNB category. For other entity types, this model showed extremely weak results. 
It was expected that training with Flair Embeddings and Transformers will not bring high recognition accuracy with corpus No.1, however, interesting results can be observed. Thus, Flair Embeddings showed decent accuracy of 0.8 for individuals with the small training dataset. 

The imbalance in the performance of different types of entities can be explained by the nature of the data, on which the original models were trained. Thus, Flair Embeddings were trained on the 1-billion words English corpus \citep{chelba_one_2013}. RoBERTa was pre-trained on the combination of five datasets containing news articles, blog entries, books, and Wikipedia articles. TARS was mainly pre-trained on datasets for text classification. Thus, the models used were not trained on domain-specific data. This can also explain the pure Transformers and TARS performance. 
The higher accuracy for the individuals category in the training with TARS can be explained by the fact, that the word 'person' is semantically more straightforward than other categories. The same could be applied to grant numbers. Furthermore, grant numbers generally have similar patterns, which can be applied to all entities of this type, that can explain a rapid improvement in F-1 score between zero-shot and few-shot models. Moreover, IND and GRNB categories showed better performance for other algorithms too, which could lie in the structure of these entities: names of individuals and grant numbers usually have undiversified patterns and in acknowledgement texts are used in a small variety of contexts. At the same time, other entity types, such as funding organisations and universities could have similar patterns and could be used in the same context. In some cases, even for human annotators, it is impossible to distinguish between university, funding body and corporation without background knowledge about the entity.

Previous works showed improvements in downstream tasks using embedding models fine-tuned for the domain used \citep{shen_sscibert_2022, beltagy_scibert_2019}. Therefore, fine-tuning the general language model on the sample of acknowledgment texts could improve the performance of the NER model for acknowledgment texts. We are planning to fine-tune BERT and Flair Embeddings (contextual string embeddings) on a sample of approx. 5 million acknowledgment texts from WoS and evaluate the performance of the NER models.   

The results of Experiment 2 generally did not show an improvement in accuracy. On the contrary, training with the three entity types deteriorated the model performance. Training without the MISC category did not show significant performance progress either. Moreover, further analysis of acknowledged entities showed that the miscellaneous category contained very inhomogeneous and partly irrelevant data, making the analysis more complicated \citep{smirnova_comprehensive_2022}.
Therefore, we assume that the model would make better predictions if the number of entity types is expanded and miscellaneous categories excluded, i.e., the MISC category could be split into the following categories: names of projects, names of conferences, names of software and dataset. Different subcategories could also be distinguished in the FUND category. 

Corpora No.2 and No.3 contain the same number of MISC and COR entities\footnote{These differences in entity distribution are caused by the peculiarities of acknowledgement information stored in WoS. As only acknowledgements with indexed funding information are stored in the database, it was difficult to find an adequate number of acknowledged entities of other types}, while in corpus 4 number of occurrences of MISC and COR entities is higher. For MISC and COR, accuracy slightly increased with corpus 4, therefore we assume that the extraction accuracy for these entities will increase with the increase of the training data. 
The situation is different for funding organizations and universities. The number of UNI and FUND entities increased evenly from corpus No.1 to corpus No.4. Nevertheless, the best result for the UNI category was achieved with corpus No.3. The poor performance of corpus No.4 could be explained by the inclusion of Indian funders. Thus, the names of many Indian funders are very similar to the entities which usually fall into the UNI category, e.g., the Department of Science and Technology or the Department of Biotechnology. This pattern is more common to the entities which fall into the UNI category. Therefore, that might make the exact extraction of UNI and FUND entities more confusing. Moreover, many Indian Universities contain the name of individuals, e.g., Rajiv Gandhi University, which can cause confusion of the UNI category with the IND category. Generally, no improvement in increasing the size of the corpus for the FUND category can be explained by the ambiguous nature of the entities which fall into the FUND category and their semantical proximity with other types of entities.
Analysis of the extracted entities showed that many entities were extracted correctly, but were assigned to the wrong category \citep{smirnova_comprehensive_2022}. Therefore, an additional classification algorithm applied to extracted entities could improve the model's performance.

\section{Conclusion}\label{sec:ret}
In this paper, we evaluated different embedding models for the task of automatic extraction and classification of acknowledged entities from acknowledgment texts\footnote{The best model can be tested via an online demo available at \url{https://colab.research.google.com/drive/1Wz4ae5c65VDWanY3Vo-fj__bFjn-loL4?usp=sharing}}. The annotation of the training corpora was the most challenging and time-consuming task of all data preparation procedures. Therefore, a semi-automated approach was used to help significantly accelerate the procedure. 

The study's main limitations were its small size and just one annotator of the training corpora. Additionally, we used acknowledgments texts collected in WoS. WoS only stores acknowledgments containing funding information, therefore there was a lack of other types of entities, such as corporations or universities in the training data.

In the present paper, we aimed to answer three questions. Thus, regarding research question 1, the few-shot and zero-shot models showed very low total recognition accuracy. At the same time, it was observed that some entities performed better than others with all algorithms and training corpora. Thus, individuals gained a good F1-score over 0.8 with zero-shot and few-shot models, as also with Flair embeddings trained with the smallest corpus. With the enlargement of the training corpora, the performance of the IND category also increased and achieved an F1-score over 0.9. The GRNB category showed an adequate F-1 score of 0.76 with the few-shot algorithm trained with the smallest corpus, following training with corpus No.2 boosts the F-1 score to over 0.9. Therefore, few-shot and zero-shot approaches were not able to identify all the defined acknowledged entity classes. 

With respect to research question 2, Flair Embeddings showed the best accuracy in training with corpus No.2 (and version 0.11) and the fastest training time compared to the other models; thus, it is recommended to further use the Flair Embeddings model for the recognition of acknowledged entities.

Exploring research question 3 we observed, that the expansion of the size of a training corpus from very small (corpus No.1) to medium size (corpus No.2) massively increased the accuracy of all training algorithms. The best-performing model (Flair Embedding) was further retrained with the two bigger corpora, but the following expansion of the training corpus did not bring further improvement. Moreover, the performance of the model slightly deteriorated. 

\subsection*{Acknowledgement}

The original work was funded by the German Center for Higher Education Research and Science Studies (DZHW) via the project "Mining Acknowledgement Texts in Web of Science (MinAck)"\footnote{\url{https://kalawinka.github.io/minack/}}. Access to the WoS data was granted via the Competence Centre for Bibliometrics\footnote{\url{https://www.bibliometrie.info/en/index.php?id=home }}. Data access was funded by BMBF (Federal Ministry of Education and Research, Germany) under grant number 01PQ17001. Nina Smirnova received funding from the German Research Foundation (DFG)  via the project "POLLUX"\footnote{\url{https://www.pollux-fid.de/about}}. The present paper is an extended version of the paper "Evaluation of Embedding Models for Automatic Extraction and Classification of Acknowledged Entities in Scientific Documents" \citep{smirnova_evaluation_2022} presented at the 3rd Workshop on Extraction and Evaluation of Knowledge Entities from Scientific Documents (EEKE2022).

\clearpage
\section{Declarations}
\subsection*{Funding and/or Conflicts of interests/Competing interests}

Philipp Mayr, the co-author of this paper, has a conflict of interest because he serves on the editorial board of the journal Scientometrics. In addition, he is a co-guest editor of the special issue on "Extraction and Evaluation of Knowledge Entities from Scientific Documents". He declares that he has nothing to do with the decision about this paper submission. 

\clearpage
\begin{appendices}
\section{Accuracy metrics by type of entity (label) for all experiments}\label{app:class_accuracy}

\begin{table}
\centering
\resizebox{\columnwidth}{!}{%
\begin{tabular}{llrlrrrrr}
\toprule
                 algorithm & corpus &  version & label &  precision &  recall &  f1-score &  support &  experiment \\
\midrule
          Flair Embeddings &   No.1 &        9 &   IND &     0.7692 &  0.8333 &    0.8000 &       12 &           1 \\
          Flair Embeddings &   No.1 &        9 &  GRNB &     0.5385 &  0.7000 &    0.6087 &       10 &           1 \\
          Flair Embeddings &   No.1 &        9 &  MISC &     0.0000 &  0.0000 &    0.0000 &        6 &           1 \\
          Flair Embeddings &   No.1 &        9 &   UNI &     0.0000 &  0.0000 &    0.0000 &        3 &           1 \\
          Flair Embeddings &   No.1 &        9 &   COR &     0.0000 &  0.0000 &    0.0000 &        1 &           1 \\
          Flair Embeddings &   No.1 &        9 &  FUND &     0.4000 &  0.4444 &    0.4211 &       18 &           1 \\
          Flair Embeddings &   No.2 &        9 &  FUND &     0.6524 &  0.7771 &    0.7093 &      157 &           1 \\
          Flair Embeddings &   No.2 &        9 &   IND &     0.9764 &  0.9831 &    0.9797 &      295 &           1 \\
          Flair Embeddings &   No.2 &        9 &  GRNB &     0.9398 &  0.9750 &    0.9571 &      160 &           1 \\
          Flair Embeddings &   No.2 &        9 &   UNI &     0.7527 &  0.7071 &    0.7292 &       99 &           1 \\
          Flair Embeddings &   No.2 &        9 &  MISC &     0.6420 &  0.6341 &    0.6380 &       82 &           1 \\
          Flair Embeddings &   No.2 &        9 &   COR &     0.8750 &  0.5833 &    0.7000 &       12 &           1 \\
         TARS (pretrained) &   No.1 &        9 &   IND &     1.0000 &  0.7500 &    0.8571 &       12 &           1 \\
         TARS (pretrained) &   No.1 &        9 &  GRNB &     0.7273 &  0.8000 &    0.7619 &       10 &           1 \\
         TARS (pretrained) &   No.1 &        9 &  MISC &     0.0000 &  0.0000 &    0.0000 &        6 &           1 \\
         TARS (pretrained) &   No.1 &        9 &   UNI &     0.0000 &  0.0000 &    0.0000 &        3 &           1 \\
         TARS (pretrained) &   No.1 &        9 &   COR &     0.0000 &  0.0000 &    0.0000 &        1 &           1 \\
         TARS (pretrained) &   No.1 &        9 &  FUND &     0.3158 &  0.3333 &    0.3243 &       18 &           1 \\
         TARS (pretrained) &   No.2 &        9 &  FUND &     0.7257 &  0.8089 &    0.7651 &      157 &           1 \\
         TARS (pretrained) &   No.2 &        9 &   IND &     0.9281 &  0.8746 &    0.9005 &      295 &           1 \\
         TARS (pretrained) &   No.2 &        9 &  GRNB &     0.8895 &  0.9563 &    0.9217 &      160 &           1 \\
         TARS (pretrained) &   No.2 &        9 &   UNI &     0.7407 &  0.6061 &    0.6667 &       99 &           1 \\
         TARS (pretrained) &   No.2 &        9 &  MISC &     0.6719 &  0.5244 &    0.5890 &       82 &           1 \\
         TARS (pretrained) &   No.2 &        9 &   COR &     0.5000 &  0.5833 &    0.5385 &       12 &           1 \\
              Transformers &   No.1 &        9 &  GRNB &     0.3000 &  0.6000 &    0.4000 &       10 &           1 \\
              Transformers &   No.1 &        9 &   IND &     0.0000 &  0.0000 &    0.0000 &       12 &           1 \\
              Transformers &   No.1 &        9 &  MISC &     0.0000 &  0.0000 &    0.0000 &        6 &           1 \\
              Transformers &   No.1 &        9 &   UNI &     0.0000 &  0.0000 &    0.0000 &        3 &           1 \\
              Transformers &   No.1 &        9 &   COR &     0.0000 &  0.0000 &    0.0000 &        1 &           1 \\
              Transformers &   No.1 &        9 &  FUND &     0.2414 &  0.3889 &    0.2979 &       18 &           1 \\
              Transformers &   No.2 &        9 &  FUND &     0.6211 &  0.7516 &    0.6801 &      157 &           1 \\
              Transformers &   No.2 &        9 &   IND &     0.9346 &  0.9695 &    0.9517 &      295 &           1 \\
              Transformers &   No.2 &        9 &  GRNB &     0.8704 &  0.8812 &    0.8758 &      160 &           1 \\
              Transformers &   No.2 &        9 &   UNI &     0.6476 &  0.6869 &    0.6667 &       99 &           1 \\
              Transformers &   No.2 &        9 &  MISC &     0.4767 &  0.5000 &    0.4881 &       82 &           1 \\
              Transformers &   No.2 &        9 &   COR &     0.7500 &  0.5000 &    0.6000 &       12 &           1 \\
  Flair Embeddings (3 Ent) &   No.2 &        9 &   IND &     0.9577 &  0.9703 &    0.9639 &      303 &           2 \\
  Flair Embeddings (3 Ent) &   No.2 &        9 &   ORG &     0.6400 &  0.6154 &    0.6275 &      208 &           2 \\
  Flair Embeddings (3 Ent) &   No.2 &        9 &  GRNB &     0.9286 &  0.9750 &    0.9512 &      160 &           2 \\
  Flair Embeddings (5 Ent) &   No.2 &        9 &   IND &     0.9764 &  0.9797 &    0.9780 &      295 &           2 \\
  Flair Embeddings (5 Ent) &   No.2 &        9 &  GRNB &     0.9345 &  0.9812 &    0.9573 &      160 &           2 \\
  Flair Embeddings (5 Ent) &   No.2 &        9 &   UNI &     0.7802 &  0.7172 &    0.7474 &       99 &           2 \\
  Flair Embeddings (5 Ent) &   No.2 &        9 &   COR &     0.7500 &  0.5000 &    0.6000 &       12 &           2 \\
  Flair Embeddings (5 Ent) &   No.2 &        9 &  FUND &     0.6722 &  0.7707 &    0.7181 &      157 &           2 \\
Flair Embeddings (RoBERTa) &   No.2 &        9 &   IND &     0.9206 &  0.9831 &    0.9508 &      295 &           2 \\
Flair Embeddings (RoBERTa) &   No.2 &        9 &  GRNB &     0.8896 &  0.9062 &    0.8978 &      160 &           2 \\
Flair Embeddings (RoBERTa) &   No.2 &        9 &   UNI &     0.5963 &  0.6566 &    0.6250 &       99 &           2 \\
Flair Embeddings (RoBERTa) &   No.2 &        9 &  MISC &     0.4135 &  0.5244 &    0.4624 &       82 &           2 \\
Flair Embeddings (RoBERTa) &   No.2 &        9 &   COR &     1.0000 &  0.5000 &    0.6667 &       12 &           2 \\
Flair Embeddings (RoBERTa) &   No.2 &        9 &  FUND &     0.6096 &  0.7261 &    0.6628 &      157 &           2 \\
          Flair Embeddings &   No.2 &       11 &  GRNB &     0.9345 &  0.9812 &    0.9573 &      160 &           3 \\
          Flair Embeddings &   No.2 &       11 &   IND &     0.9797 &  0.9831 &    0.9814 &      295 &           3 \\
          Flair Embeddings &   No.2 &       11 &  FUND &     0.7027 &  0.8280 &    0.7602 &      157 &           3 \\
          Flair Embeddings &   No.2 &       11 &   UNI &     0.7684 &  0.7374 &    0.7526 &       99 &           3 \\
          Flair Embeddings &   No.2 &       11 &  MISC &     0.6543 &  0.6463 &    0.6503 &       82 &           3 \\
          Flair Embeddings &   No.2 &       11 &   COR &     0.7500 &  0.5000 &    0.6000 &       12 &           3 \\
          Flair Embeddings &   No.3 &       11 &   UNI &     0.8000 &  0.7273 &    0.7619 &       99 &           3 \\
          Flair Embeddings &   No.3 &       11 &   IND &     0.9731 &  0.9797 &    0.9764 &      295 &           3 \\
          Flair Embeddings &   No.3 &       11 &  GRNB &     0.9281 &  0.9688 &    0.9480 &      160 &           3 \\
          Flair Embeddings &   No.3 &       11 &   COR &     0.7500 &  0.5000 &    0.6000 &       12 &           3 \\
          Flair Embeddings &   No.3 &       11 &  MISC &     0.6571 &  0.5610 &    0.6053 &       82 &           3 \\
          Flair Embeddings &   No.3 &       11 &  FUND &     0.6757 &  0.7962 &    0.7310 &      157 &           3 \\
          Flair Embeddings &   No.4 &       11 &  MISC &     0.7424 &  0.5976 &    0.6622 &       82 &           3 \\
          Flair Embeddings &   No.4 &       11 &   COR &     0.8571 &  0.5000 &    0.6316 &       12 &           3 \\
          Flair Embeddings &   No.4 &       11 &   UNI &     0.7753 &  0.6970 &    0.7340 &       99 &           3 \\
          Flair Embeddings &   No.4 &       11 &   IND &     0.9698 &  0.9797 &    0.9747 &      295 &           3 \\
          Flair Embeddings &   No.4 &       11 &  FUND &     0.6823 &  0.8344 &    0.7507 &      157 &           3 \\
          Flair Embeddings &   No.4 &       11 &  GRNB &     0.9162 &  0.9563 &    0.9358 &      160 &           3 \\
\bottomrule
\end{tabular}
}
\caption{\label{tab:app_a} Accuracy metrics by type of entity (label) for all experiments. Rows are sorted by experiment number and algorithm.}
\end{table}

\section{Overall accuracy for all experiments}\label{app:accuracy}

\begin{table}
\centering
\resizebox{\columnwidth}{!}{%
\begin{tabular}{llrrr}
\toprule
                        algorithm & corpus &  version &  accuracy &  experiment \\
\midrule
                 Flair Embeddings &   No.2 &        9 &    0.7702 &           1 \\
                 Flair Embeddings &   No.1 &        9 &    0.3472 &           1 \\
                TARS (pretrained) &   No.2 &        9 &    0.7113 &           1 \\
                TARS (pretrained) &   No.1 &        9 &    0.3485 &           1 \\
                     Transformers &   No.2 &        9 &    0.6783 &           1 \\
                     Transformers &   No.1 &        9 &    0.1477 &           1 \\
Flair Embeddings (3 Entity Types) &   No.2 &        9 &    0.7536 &           2 \\
Flair Embeddings (5 Entity Types) &   No.2 &        9 &    0.7990 &           2 \\
       Flair Embeddings + RoBERTa &   No.2 &        9 &    0.6697 &           2 \\
                 Flair Embeddings &   No.2 &       11 &    0.7869 &           3 \\
                 Flair Embeddings &   No.4 &       11 &    0.7814 &           3 \\
                 Flair Embeddings &   No.3 &       11 &    0.7691 &           3 \\
\bottomrule
\end{tabular}
}
\caption{\label{tab:app_b} Overall accuracy for all experiments. Rows are sorted by experiment number and algorithm.}
\end{table}

\end{appendices}

\clearpage
\typeout{} 
\bibliography{sn-bibliography}% common bib file

%\bibliographystyle{sn-apacite}


\end{document}